# Prolonged spontaneous emission and dephasing of localized excitons in air-bridged carbon nanotubes

*Ibrahim Sarpkaya[1], Zhengyi Zhang[2], William Walden-Newman[1], Xuesi Wang[2], James Hone[2], Chee Wei Wong[2], and Stefan Strauf[1*]*

[1] Department of Physics, Stevens Institute of Technology, Hoboken, NJ 07030, USA

[2] Department of Mechanical Engineering, Columbia University, New York, NY 10027, USA

*Address correspondence to: strauf@stevens.edu

The bright exciton emission of semiconducting carbon nanotubes is appealing for optoelectronic and quantum photonic devices as well as fundamental studies of light-matter interaction in one-dimensional nanostructures. The photophysics of excitons in carbon nanotubes till date is however largely affected by extrinsic effects of the dielectric environment. Here we carried out time-resolved photoluminescence measurements over 14 orders of magnitude for ultra-clean carbon nanotubes bridging an airgap over pillar posts. Our measurements demonstrate a new regime of intrinsic exciton photophysics with prolonged spontaneous emission times up to $T_1$=18 ns, between one to two orders of magnitude better than prior measurements and in agreement with values first hypothesized by theorists about a decade ago. Further we establish for the first time the exciton decoherence times of individual nanotubes in the time-domain and found four-fold prolonged values up to $T_2$=2.1 ps compared to ensemble measurements. These first observations motivate new discussions about the magnitude of the intrinsic dephasing mechanism while the prolonged exciton dynamics is promising for device applications.

The photophysics of excitons in single-walled carbon nanotubes (SWCNTs) has been the subject of much recent interest [1]. An appealing property for optoelectronic device applications as well as fundamental studies is the weakly-screened Coulomb interaction giving rise to exciton binding energies up to 400 meV [2], which are about an order of magnitude larger than for excitons in II-VI or III-V semiconductor quantum dots. Unlike quantum dots, the photoluminescence (PL) quantum yield in one-dimensional SWCNTs is rather low [3]. It is however increasingly understood that this is caused by *extrinsic* nonradiative recombination (NR) of highly mobile excitons exploring defects and dopants along the tube [4-11], and to a lesser extend due to *intrinsic* exciton dark states affecting PL yields only below 50K [12]. There is also increasing evidence that excitons localize randomly along the tube and form quantum dot like states [13], as is indicative from near-field scanning experiments [14] and the observation of photon antibunching at low temperatures [15,16]. As a result of these extrinsic effects, dynamical properties of excitons are affected over many timescales in experiments giving rise to PL blinking and spectral diffusion (SD) on millisecond to minute time scales [16,17], spontaneous emission lifetimes ($T_1$) of 20 to 200 ps [5, 7, 12, 15, 18] and ultrafast exciton decoherence or dephasing time ($T_2$) of up to about 500 fs till date [19,20]. So far, exciton dephasing has only been investigated in ensemble studies which are likely affected by surfactants, inter-tube and tube-substrate coupling effects.

The question arises to what extend these experiments probe the true intrinsic properties of the exciton dynamics and how one could reach deeper into the intrinsic regime. In particular, theorists predicted about a decade ago $T_1$ times of excitons of about 10 ns [21,22] while measurements have lifetimes varying from less than 20 to 200 ps. Recently, Perebeinos and Avouris proposed a mechanism which attributes this discrepancy to residual *p*-type doping opening up an efficient NR recombination channel via phonon-assisted indirect exciton ionization [23]. This implies that the ps-fast $T_1$ times are extrinsic and that intrinsic exciton properties could be expected when doping is reduced. To this end it is appealing to study SWCNTs that are isolated from their environment by suspending them over an air-gap done by one of the authors [24], an approach that can eliminate blinking and SD on millisecond to minute time scales [25]. Recent studies of air-bridged SWCNTs estimate $T_2$ indirectly using the spectral linewidth [26]. However, since the linewidth can be in principle affected by SD, it is not a reliable measure for exciton dephasing and only a lower temporal bound. Direct measurements of exciton dephasing in the time domain have not yet been carried out for individual SWCNTs.

Here we report time-resolved photoluminescence measurements over 14 orders of magnitude for individual ultra-clean SWCNTs bridging an airgap over pillar posts. Our cryogenic single nanotube measurements demonstrate a new regime of intrinsic localized exciton photophysics with narrow spectral linewidth down to 200 μeV and prolonged spontaneous emission times up to 18 ns, approaching into the intrinsic regime. In addition, we establish for the first time the exciton dephasing times for individual SWCNTs in the time-domain and found four-fold prolonged values up to $T_2$=2.1 ps when compared to ensemble measurements. Furthermore, our experiments demonstrate that the spectral linewidth is not a reliable measure for the exciton dephasing time in environment affected SWCNTs of prior studies.

**Results**

**Exciton emission from ultra-clean air-bridged SWCNTs.**

In order to investigate the exciton photophysics without distortion from a dielectric surrounding or surfactants we have grown SWCNTs in such a way that they bridge an air gap between pillar posts. **Figure 1a** shows a corresponding scanning electron micrograph (SEM) image of an individual SWCNT with a length of about 2 μm. The CNT nucleation starts predominantly from a Co-layer deposited on top of the pillars acting as a catalyst. The growth direction follows the gas-flow stream until the other end is reached [13, 27]. More details are given in the Methods section. From scanning-electron microscope (SEM) scans of about 1000 pillar post pairs we determine a success rate of 15% that an individual SWCNT is fully elevated. In some cases the SWCNT would grow from the top diagonal down to the substrate but still being elevated along its entire length, while in most cases the location is simply empty. As a result, spatial scans in micro-PL reveal single narrow emission lines in the spectrum in case of semiconducting SWCNTs. In addition, the emission lines display pronounced intensity dependence upon rotation of the linear incident laser polarization, indicating that the optical emission stems from individual SWCNTs [16,17]. The spectrum in **Figure 1b** demonstrates single Lorentzian lineshapes of the $E_{11}$ exciton recombination and the corresponding spectral trajectory in **Figure 1c** shows that the emission is stable with no significant spectral diffusion or blinking at 200 ms timing resolution. The 890 nm emission wavelength is attributed to SWCNTs with (6,4) chirality [28]. While we find SWCNTs with various chiralities along the wafer we focus here on the (6,4) tubes to avoid any potential variations in the exciton photophysics due to differences in the tube diameter.

Interestingly, we do not find any evidence of asymmetric exciton lineshapes with linewidth of about 3.5 meV as previously reported for surfactant embedded SWCNTs grown by the CoMoCat technique, which were attributed to an intrinsic dephasing mechanism of excitons and acoustic phonons [29]. Such an asymmetry was also reported for air-bridged SWCNTs displaying rather large linewidth of 10 meV and attributed to the Van Hove singularities in the density of states [30]. The relatively large linewidth of 10 meV in the absence of a substrate as found in Ref. 30 is also indicative that even air-bridged SWCNTs can be affected by the environment, possibly due to residual unintentional doping.

To shine more light on this issue we targeted the growth of ultra-clean SWCNTs with the idea that shorter growth time should result in less residual amorphous carbon in the SWCNT vicinity. In **Figure 2** we compare the spectral linewidth as a function of pump power for surfactant dispersed SWCNTs of (6,4) chirality embedded in a polystyrene matrix (blue triangles), with air-bridged SWCNTs grown for 10 min (black squares), and air-bridged SWCNTs grown for only 2 min (red circles). Compared to surfactant dispersed SWCNTs with linewidth up to 10 meV the air-bridged SWCNTs from the 10 min growth are significantly narrower. Best values down to 220 µeV are found in the 2 min growth run. Since it is well known that charge fluctuations from residual doping can give rise to spectral diffusion resulting in linewidth broadening [16, 17], one can conclude that observation of ultra-narrow spectral emission lines in the 2 min growth is caused by ultra-clean SWCNTs.

The narrow and symmetric lineshape implies that the proposed acoustic phonon dephasing mechanism giving rise to asymmetric and broader lineshapes is either not intrinsic, i.e. substrate, dopant, and/or surfactant induced, or not relevant at a comparable magnitude when SWCNTs are suspended in air. In particular, our observation of a spectral linewidth of 220 µeV implies a lower limit for the dephasing time $T_2$ of 6 ps, which is about an order of magnitude longer than previously assumed. In addition, one can expect that the spontaneous emission lifetime $T_1$ becomes significantly longer if the PAIEI nonradiative decay channel is effectively suppressed in ultra-clean SWCNTs.

**Spectral diffusion via photon correlation spectroscopy.**

To further investigate the photophysics of excitons in these air-bridged SWCNTs and demonstrate prolonged $T_1$ and $T_2$ times as the signature of the intrinsic regime we carried out time-resolved measurements over 14 orders of magnitude from minutes down to the sub-ps regime. The emission energy and intensity of the exciton recombination of an individual quantum emitter is well known to strongly fluctuate with time due to spectral diffusion (SD) and blinking, which can give rise to significant linewidth broadening and low quantum efficiency. These detrimental quantum fluctuations are however to a large part not intrinsic and can be strongly suppressed, either by embedding SWCNTs into a polystyrene matrix as we have shown recently [16,17], or by dispersing the SWCNT in air [25]. Our data from the air-bridged SWCNTs shown in **Figure 1c** confirm these earlier findings, i.e. there are no pronounced effects of SD or blinking even for the 10 min growth samples when streaming photons for 5 minutes at a timing resolution of 100 ms. However, such studies of SD and blinking based on direct streaming of the photon flux with CCD cameras or single photon avalanche photodiodes (APDs) are limited to time scales of ms or μs at best. This raises the question if the residual linewidth of air suspended SWCNTs is affected by SD on faster time scales which are not resolved in these experiments.

In order to investigate such contributions of SD down to the ns time scale, we describe in the following an experiment based on photon correlation spectroscopy [16, 31] applied here for the first time to air suspended SWCNTs. **Figure 3a** illustrates the measurement principle: We assume that the measured PL spectrum (blue) is composed of intrinsically much narrower spectral lines that rapidly fluctuate around a center wavelength due to SD. By sending light from a specific sub-region of the PL spectrum selected by a bandpass filter (pink shaded area) towards the single photon correlation setup one creates a conditional detection probability that a second photon will be detected from the same spectral region at a later time. This detection probability eventually decreases with time if the exciton recombination energy shifts out of the spectral detection window due to SD, i.e. the optical Stark shift generated by fluctuating charges. As a result, the signature of SD in the second–order photon correlation function $g^2(\tau)$ is a bunching signal centered at zero delay time decaying with the characteristic SD time.

For comparison we first illustrate in **Figure 3b** a SWCNT which has been dispersed in a surfactant and embedded in a polymer matrix (see Methods). This SWCNT displays pronounced bunching when filtered with a 10 nm bandpass centered over the right half of the emission spectrum. The bunching disappears when filtered with a 40 nm wide filter covering the entire PL spectrum as expected, since in this case no conditional probability is created at the detector. The SD decay time can be quantified using the relation $g^2(\tau) = \left[1 + \left(\frac{\gamma_d}{\gamma_{L,R}} - 1\right) e^{(-\gamma_d \tau)}\right]$, where $\gamma_{L,R}$ is the crossover rate for shifting in and out of the filter range which determines the peak height, and $\gamma_d$ is the SD rate which determines the SD time $\tau_d = 1/\gamma_d$. From the solid red curve fit of $g^2(\tau)$ in **Figure 3b** we find $\tau_d$ = 3.7± 0.4 ns. These SD times vary as a function of pump power as well as dielectric environment and are most pronounced in uncapped SWCNTs touching a SiO$_2$ substrate, as further detailed in Ref. 16.

In contrast, **Figure 3c** demonstrates that such a bunching signature is completely absent in air-bridged SWCNTs from the 10 min growth. The experiment was carried out at a pump power of 1.8 mW were the particular SWCNT displays a spectral linewidth of 2.2 meV (1.4 nm), which was filtered with a 1 nm bandpass filter positioned over the right half of the emission spectrum. By recording the transmitted spectrum in each arm before APDs we confirmed that a conditional probability is created (see supplemental Figure). We tested more than 20 air-bridged SWCNTs and never found bunching while every surfactant dispersed SWCNT did reliably produce a bunching signature. As a result, the photon correlation experiments reveal that there are no contributions of SD in air-bridged SWCNTs down to the system resolution limit of about 200 ps. One reason why the correlation trace for the air-bridged tubes does not show bunching from residual SD could be that it is mask by photon antibunching from localized quantum dot like excitons. Measurements with a 10 nm broad filter before detectors can exclude this as an explanation (see Supporting Online information). Therefore, our measurements show that that either there is no SD, or, which is more likely in light of our findings in **Figure 5e**, that residual SD contributes to the linewidth, but is much faster than what can be resolved with the correlation technique.

**Prolonged spontaneous emission lifetimes.**

With this clear advantage of air-bridged SWCNTs compared to surfactant dispersed SWCNTs one can now examine the intrinsic properties of the exciton photophysics. In particular, we expect that the spontaneous emission time $T_1$ approaches the intrinsic limit of several ns, which was theoretically predicted by Perebeinos et al. for the case of moderate exciton localization and when phonon-assisted indirect exciton ionization (PAIEI) can be neglected [23]. The PAIEI mechanism is caused by unintentional doping pinning the Fermi level in the valence band, which opens up a very effective nonradiative decay channel via phonon-assisted intraband electron-hole pair generation. In this sense, the absence of SD down to sub-ns time scales in **Figure 3c** as well as the narrow spectral linewidth in **Figure 2** is very promising, implying a lower residual dopant concentration and thus suppressed PAIEI as compared to surfactant-dispersed SWCNTs.

The dynamics of the exciton recombination was recorded by time correlated single photon counting (TCSPC) as shown in **Figure 4**. For our comparison group of surfactant-dispersed SWCNTs we find in all cases nearly resolution limited $T_1$ times, as shown in **Figure 4a** (blue triangles). After deconvolution of the system response function (210 ps) an upper bound for the $T_1$ times of about 30±20 ps is estimated.

In stark contrast all investigated air-bridged SWCNTs display a pronounced bi-exponential decay behavior (see **Figure 4a**), with a moderately long $T_1$ time of 30-200 ps and a second ultra-long decay time of up to 10 ns for the 10 min growth, and up to 18 ns for the 2 min growth. Statistical results recorded from 20 individual air-bridged SWCNTs are shown in **Figure 4b**. Average values of the 10 min growth are found to be $T_1$= 5.7 ns and for the 2 min growth $T_1$=9 ns, as indicated by the horizontal lines. As a result for the three groups of SWCNTs, narrower linewidth as shown in **Figure 2** correlates with longer $T_1$ times as shown in **Figure 4**. This strongly indicates that the effective removal of unintentional doping gradually removes the PAIEI decay channel, giving rise to the observed ns decay component.

The pronounced variation of $T_1$ times in the air-bridged SWCNTs from 4 to 18 ns and the contribution of a 30 to 200 ps fast component are attributed to the varying degree of exciton localization and any residual unintentional doping. From transport measurements of comparable SWCNTs grown in the same CVD furnace it is known that the electron mean free path varies between 400 nm and 10 μm, recorded at 10 K where one can neglect contributions from acoustic phonons to transport [32]. These variations in the electron mean free path have been attributed to the static disorder caused by residual impurities and they correlate with the number of defects determined from scanning gate microscopy [32]. Therefore, if a particular section along the tube is less affected by residual doping, the PAIEI mechanism responsible for the 30 to 200 ps fast nonradiative decay is effectively suppressed, while excitons in other sections of the SWCNT can still recombine fast. In the clean sections and absence of PAIEI the $T_1$ time is only adversely affected by intrinsic multiphonon decay, which strongly varies in presence of exciton localization. Calculations of $T_1$ times for SWCNT with diameter of 1 nm predict $T_1$ values of 100 ns in the absence of exciton localization, which speeds up to about 0.5 ns for strong exciton localization along the tube [23]. Therefore, our measured $T_1$ values of 4 to 18 ns reveal the onset of a remarkable new regime of intrinsic localized exciton recombination approaching spontaneous emission times predicted about a decade ago by theorists [21,22].

**Prolonged exciton dephasing.**

Finally, we present exciton dephasing measurements for the first time for individual SWCNTs in the time domain by recording the first-order autocorrelation function $g^1(\tau)$ with a Michelson interferometer. **Figure 5a** shows the interferogram of light from the $E_{11}$ exciton recombination. The right panel illustrates the high fringe visibility with values up to 97% achieved by a delay line with retro mirrors (**Figure 5b**) and a stepping resolution of 50 nm. The fringe visibility $V(\tau)$ was estimated using the relation $V(\tau)=(I_{max}(\tau)-I_{min}(\tau))/(I_{max}(\tau)+I_{min}(\tau))$, where $I_{max}(\tau)$ and $I_{min}(\tau)$ are the fringe maxima and minima determined from fits to the fringe pattern at various delay times. **Figure 5c** shows $V(\tau)$ as a function of delay time which fits monoexponentially with decay time $T_2 = 1.4\pm0.2$ ps. At a fringe visibility at or below 35% the data become too noisy to be reliable, constituting the detection limit, as indicated by the black shading. Dephasing times determined in this way are plotted in **Figure 5d** for all three cases. While surfactant dispersed, polymer embedded SWCNTs show on average the fastest dephasing with 830 fs, air-bridged SWCNTs from the 10 min growth display an average $T_2=1.22$ ps, while the cleaner 2 min growth yields an average of $T_2=1.38$ ps with a longest value of $T_2=2.1$ ps, which corresponds to a four-fold prolonged exciton dephasing time when compared to ensemble measurements [19,20]. Our linewidth measurements down to 220 µeV at lowest pump power imply furthermore a lower limit of 6 ps for the $T_2$ time of the air-bridged SWCNTs from the 2 min growth. The interferometric measurements at the level of individual SWCNT require however larger pump powers resulting in broader linewidth (see **Figure 2**), limiting the experimental accessible range for $T_2$ measurements to the high pump powers.

**Discussions**

It is interesting to compare exciton dephasing times estimated by the direct time-domain approach with the indirect frequency-domain measurements, i.e. the spectral linewidth $\Gamma$ of the PL emission. If there is no other physical mechanism affecting the spectral linewidth, such as spectral diffusion, than both values from time and frequency domain experiment should match and follow the well-known relation $\Gamma = \hbar/T_1 + 2\hbar/T_2$. For this purpose **Figure 5e** plots the difference in energy between the measured spectral PL linewidth and the spectral linewidth which can be estimated with the above relation from the measured $T_2$ values from **Figure 5d.** In estimating the linewidth we corrected for possible broadening from the $T_1$ process, which is however a rather small correction, such that $\Gamma$ is dominated by the dephasing time alone (see supporting online material). The deviations for polymer embedded SWCNTs up to 6 meV and air-bridged SWCNTs from the 10 min growth up to 3.7 meV in **Figure 5e** are striking and imply that the exciton dephasing time in SWCNTs cannot be accurately determined from a time-integrated linewidth study alone, as was previously suggested [26]. In contrast, our ultra-clean samples from the 2 min growth show little to no extra broadening in Figure 5e (red dots), thus following the standard relation $\Gamma = \hbar/T_1 + 2\hbar/T_2$. This is another clear indication that emission from the samples of the 2 min growth approaches the intrinsic regime.

We furthermore note that the measured lifetimes of $T_1$= 4 to 18 ns are indicative of *localized* exciton emission since calculations of $T_1$ times for SWCNT with diameter of 1 nm predict $T_1$ values of 100 ns in the complete absence of exciton localization, which speeds up to about 0.5 ns for strong exciton localization along the tube [23]. One would thus expect that the localization can result in quantum-dot like states along the nanotube, which should give rise to nonclassical light emission in form of photon antibunching. While we do find pronounced signatures of photon antibunching under pulsed excitation in several polymer embedded SWCNTs as well as air-bridged SWCNTs (see Supplementary Figures S5 and S6), we find also many air-bridged SWCNTs which do not display photon antibunching despite their narrow linewidth and prolonged $T_1$ and $T_2$ times. Since SWCNTs can possess several localization sites along the tube length, and since the optical excitation spatially averages over them, the photon antibunching signatures can vanish in the limit of several quantum emitters. On the other hand all air-bridged SWCNTs display contributions of prolonged $T_1$ times, which have been predicted about a decade ago by theorists [21,22], and are the hallmark of the intrinsic exciton recombination.

Finally, we discuss the underlying dephasing mechanism in light of our new findings. It is well known that optical phonons dominate the exciton dephasing above a lattice temperatures of 180 K [20], while at our cryogenic measurement temperatures of 9 K exciton dephasing is dominated by one-phonon acoustic scattering with a linear bath temperature dependence [19,20,26,29]. Other processes such as collision induced broadening from elastic exciton-exciton scattering and exciton-exciton annihilation provide an additional contribution to the exciton dephasing, which is however rather small at intermediate pump powers were our experiments are carried out, while values reach up to 100 μeV in the high pump power saturation regime [33]. A recent theory also predicts suppression of exciton-electron scattering in doped SWCNTs for the optically active singlet exciton at low temperatures with contributions to dephasing of 100 μeV or less at 50 K and high doping (0.3 electrons/nm), which vanishes at low doping levels [34].

The merit of our finding is that the magnitude of the exciton acoustic-phonon scattering is sizably smaller than concluded in prior reported efforts. This is evident from both the remarkably narrow linewidth (Figure 2) and the prolonged dephasing times (Figure 5), which we observed for the first time in our experiments.

As an example of prior efforts, Ref. 29 finds asymmetric spectral lineshapes with 3.5 meV linewidth ($T_2$ about 400 fs) and applies the independent Boson model to fit the lineshape based on the magnitude of the exciton-phonon matrix elements $g_j(q)$, which is a product of the exciton form factor $F(q)$ describing coupling in momentum space and the deformation potential coupling $G_s(q)$ of the stretching mode. Agreement between theory and experiment has been found using a deformation potential value for the stretching mode of $D_s$ =12-14 eV and an exciton wavefunction envelope of about 10 nm. It is however also clear that these parameters can depend strongly on the nanotube environment. On one hand, in $F(q) = \exp(-q^2\sigma^2/4)$, the exciton confinement length $\sigma$ depends on the underlying degree of exciton localization, which can vary from tube to tube and in different environments [35]. On the other hand, $G_s(q)$ depends linearly on $D_s$ and inversely on the square root of the SWCNT length. The magnitude of the deformation potential coupling is likely smaller in air-bridged SWCNT, since for a substrate supported SWCNT the effective deformation potential can be additionally affected by substrate phonons. This is for example the case in graphene on $SiO_2$ showing strongly reduced carrier mobility due to scattering with interface phonons, while air-suspended graphene displays an order of magnitude larger mobility.

Besides acoustic phonons, disorder and/or structural defects contribute additionally to the exciton dephasing. Ab initio calculations of (6,4) SWCNTs predict that Stone-Wales or 7557 type defects can introduce a broad range of disorder modes resulting in fast exciton dephasing down to 10 fs at room temperature [36]. It is thus likely that amorphous carbon or graphitic residue decorating the SWCNT can also contribute to the exciton dephasing, in particular since the weak Coulomb screening allows the exciton wavefunction to penetrate into the tube vicinity. Our remarkably narrow spectral linewidth and the prolonged dephasing time for the air-bridged SWCNTs from the 2 min growth suggest that they have significantly lower disorder and defects than prior efforts.

As a result, our data show that the intrinsic exciton acoustic phonon coupling in ultra-clean air-bridged SWCNTs is significantly weaker than previously believed, while the presence of disorder from residual amorphous carbon or defects can give rise to spectral diffusion and faster exciton dephasing times, even in some of the air-bridged SWCNTs.

**Conclusion**

We carried out time-resolved photoluminescence measurements over 14 orders of magnitude for individual SWCNTs bridging an airgap over pillar posts. Our measurements demonstrate a new regime of intrinsic localized exciton photophysics in ultra-clean SWCNTs displaying prolonged spontaneous emission times up to $T_1$=18 ns, which have been predicted by theorists about a decade ago. Longer lifetimes are attributed to an effective suppression of phonon-assisted indirect exciton dissociation, the PAIEI mechanism, which can be caused by unintentional doping. This is in agreement with photon correlation experiments demonstrating that surfactant dispersed SWCNTs display spectral diffusion while air-bridged SWCNTs are free of spectral diffusion down to the detection limit, implying reduced background doping.

Furthermore we measured for the first time the exciton dephasing times for individual SWCNTs in the time-domain and found four-fold prolonged values up to $T_2$=2.1 ps compared to ensemble measurements. Our experiments demonstrate that the spectral linewidth is not a reliable measure for the exciton dephasing time in prior SWCNTs, unless ultra-clean SWCNTs reach deeply into the intrinsic regime. The ultra-narrow linewidth and prolonged dephasing times suggest that the matrix element for exciton acoustic-phonon coupling has been largely overestimated in previous work. While a longer dephasing is of importance for applications in quantum optics, the prolonged exciton dynamics is promising towards efficient optoelectronic and quantum photonic devices based on ultra-clean SWCNTs.

**Acknowledgement.** The authors acknowledge discussions with Tony F. Heinz and Vasili Perebeinos. S.S. acknowledges financial support by the National Science Foundation (NSF), CAREER award ECCS-1053537. X.W., Z.Z., J.H. and C.W.W acknowledge support on the materials processing from the Center for Re-Defining Photovoltaic Efficiency Through Molecule Scale Control, an Energy Frontier Research Center funded by the U.S. Department of Energy, Office of Science, Office of Basic Energy Sciences under Award Number DE-SC0001085. J.H. and Z.Z. acknowledge support from NSF DMR-1006533. X.W. and C.W.W. acknowledge support from NSF (IGERT award DGE-1069240 and CAREER award ECCS-0747787). This research effort used microscope resources partially funded by NSF through Grant DMR-0922522. Sample fabrication was carried out in part at the Center for Functional Nanomaterials, Brookhaven National Laboratory, which is supported by the U.S. Department of Energy, Office of Basic Energy Sciences, under Contract No. DE-AC02-98CH10886.

**Author contributions.** S.S. and I.S. conceived and designed the experiments. Z.Z. and X.W. fabricated the air-bridged SWCNTs. I.S., W.W., and S.S. carried out the optical experiments and analyzed the data. I.S. and S.S. co-wrote the manuscript. C.W.W. and J.H. supervised the sample fabrication and S.S. supervised the optical measurements. All authors contributed to scientific discussions.

**Competing financial interests.** The authors declare no competing financial interests.

**METHODS**

**Sample preparation**. In order to create elevated SWCNTs bridging an air gap, a 1 nm thin layer of Co was deposited by electron-beam evaporation onto a $Si/SiO_2$ wafer. Pillar pairs of 3 μm spacing and lateral walls of 5 μm spacing were defined lithographically and transferred by reactive-ion etching into the $Si/SiO_2/Co$ substrate for about 2.5 μm. Single–walled carbon nanotubes were grown by ambient chemical vapor deposition (CVD) method using a modified fast-heating process with ethanol as feedstock [37,38]. Growth was carried out at a temperature of 900º C in an $Ar/H_2$ gas mixture at typical EtOH flow rates of 80 sccm. The pillar pairs were spaced out 30 μm in order to avoid any unwanted emission from neighboring SWCNTs in the optical measurements. The growth time was varied between 2-10 min.

For comparison studies we also used commercial SWCNTs grown by the CoMoCat technique which were embedded in polystyrene on top of a gold mirror to enhance outcoupling. These samples were fabricated using an ebeam evaporator to coat a 100 nm layer of Au on top of a standard p++ type Si wafer with a 90 nm $SiO_2$ layer. This was followed by a 160 nm layer of spin-coated polystyrene. Commercial CoMoCat SWCNTs were prepared with bath sonication for 1 hour in a vial containing 0.4 wt % sodium dodecylsulphate (SDS) solution. The product was poured through a 5 μm filter to form a concentration of 0.2 mg/mL, and deposited directly onto the first layer of polymer, and then covered with a second 160 nm layer of polymer. Finally, the samples were baked at 95 °C for several hours before cryogenic measurements to remove moisture acting as charge trap states [39]. This approach effectively removes spectral diffusion and blinking as we recently demonstrated [16,17].

**Micro-photoluminescence setup.** Measurements of micro-photoluminescence (μ-PL) were taken inside a liquid helium cryostat with a 9 K base temperature. Samples were excited with a laser diode operating at 780 nm in continuous wave or pulsed mode (80 MHz repetition rate and 100 ps pulse length). A laser spot size of about 1.5 micron was achieved using a microscope objective with numerical aperture of 0.55. The relative position between sample and laser spot was adjusted with a piezo-electric *xyz*-actuator mounted directly onto the cold finger of the cryostat. Spectral emission from the sample was dispersed using a 0.75 m focal length spectrometer and imaged by a liquid nitrogen cooled CCD camera. Laser stray light was rejected combining a 780 nm notch filter and 800 nm high-pass filter. To enhance the exciton emission the laser polarization was rotated with a half-wave plate with respect to the axis of the SWCNTs. Measurements of spectral trajectory over time were obtained by continuously recording the spectra every 200 ms with a silicon CCD camera.

**Photon correlation measurements ($T_1$, $T_2$, SD).** The second-order photon correlation function $g^2(\tau)$ was recorded by sending the PL emission through narrow bandpass filters and onto a Hanbury-Brown and Twiss setup consisting of a fiber-coupled 50/50 beam splitter connected to two single photon counting avalanche photodiodes (silicon APD). A linear polarizer was used in the collection path for photon antibunching measurements and narrow bandpass filters for the SD measurements. Coincidence counts were time stamped and analyzed with a high resolution timing module. To determine spontaneous emission lifetimes ($T_1$) the 50/50 beam splitter was removed and the first-order photon correlation function $g^1(\tau)$ was recorded by sending light from the sample to the first detector (start) and an electronic trigger from the 780 nm laser diode to the stop channel on the timing module. The system response function was measured sending scattered laser light from the sample surface to the start APD.

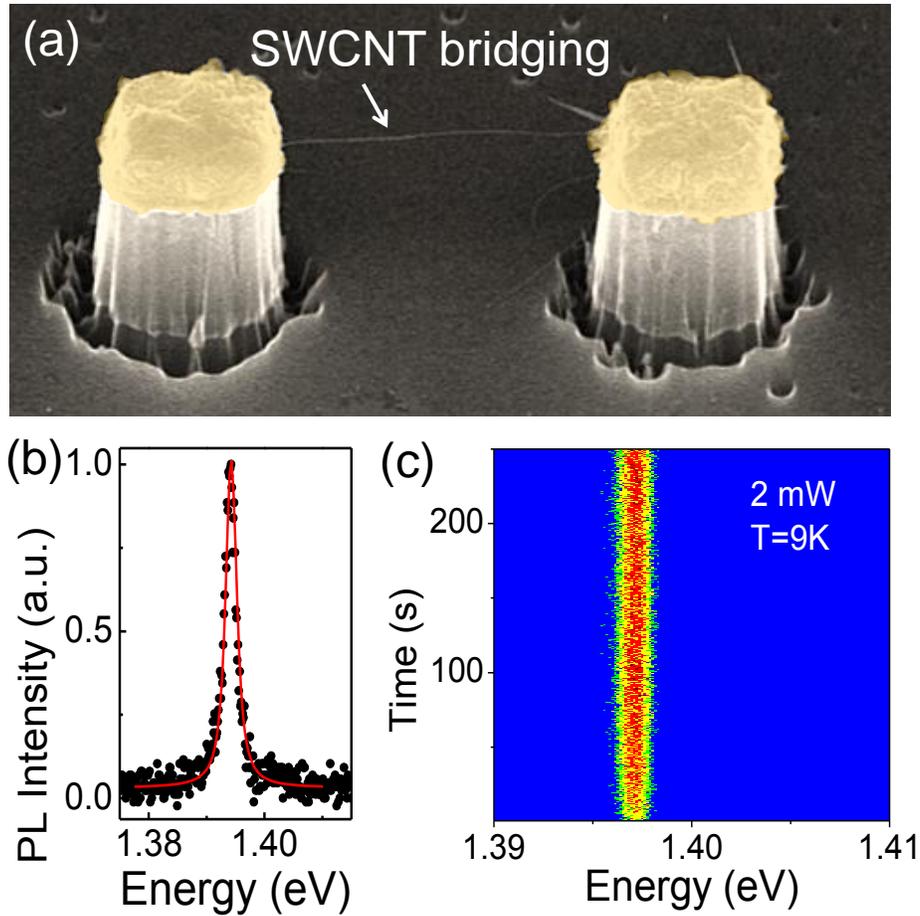

**Figure 1. Optical emission from individual air-bridged SWCNTs.** (a) Scanning electron microscope image of an individual SWCNT bridging two pillar posts which are separated 3 μm from center to center. The directional growth starts from the Co metal catalyst located on top of the pillars (shaded yellow). (b) Photoluminescence spectrum of an individual and pristine SWCNT with (6,4) chirality suspended in air. Symmetric lineshapes are observed at 9 K and pump powers of 50 μW. Solid red line is a Lorentzian fit. (c) Spectral trajectory of the PL emission recorded with 200 ms timing resolution showing no significant spectral diffusion and blinking at this timing resolution.

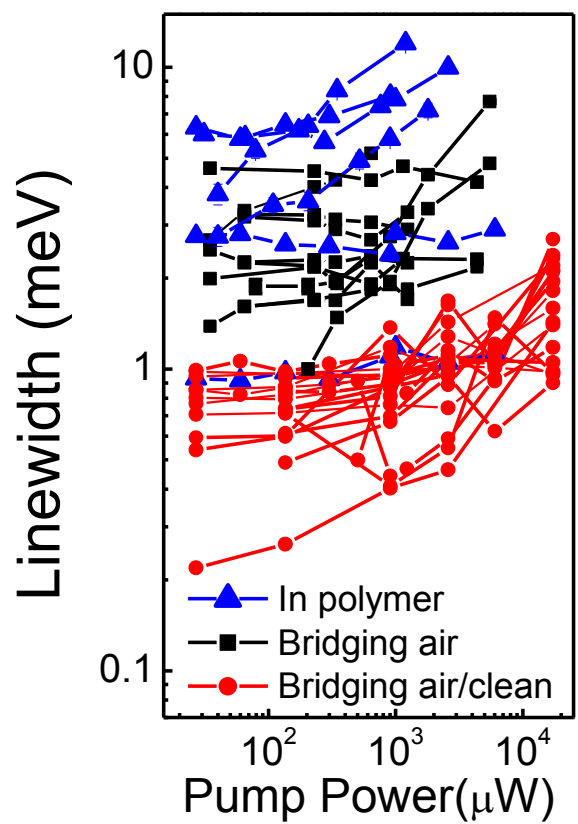

**Figure 2. Spectral linewidth study**. Linewidth versus pump power for surfactant dispersed and polymer embedded SWCNTs (blue triangles), air-bridged SWCNTs grown for 10 min (black squares), and "cleaner" air-bridged SWCNTs grown for only 2 min (red circles). All data are corrected for the spectrometer response function by fitting a Voigt function to the time integrated spectrum, i.e. a convolution of a single Lorentzian describing the exciton recombination and a Gaussian describing the system broadening. Data are recorded at 9 K.

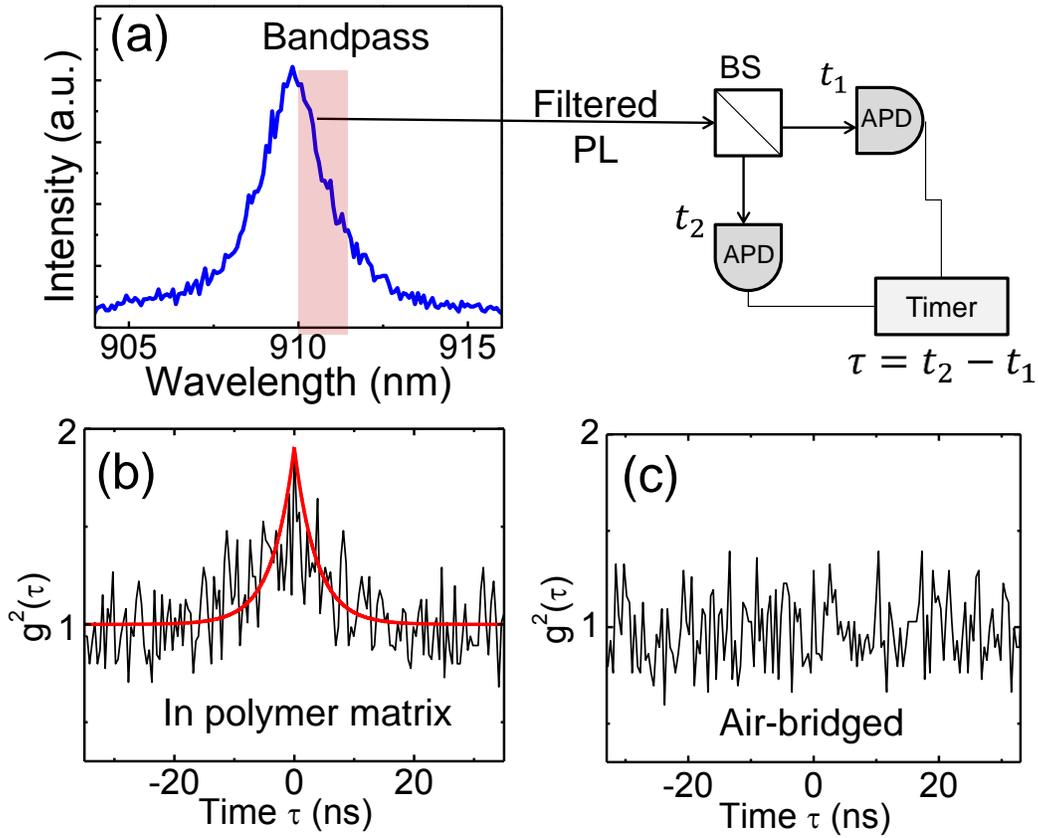

**Figure 3. Spectral diffusion studies using photon correlations measurements.** (a) Plot of PL emission spectrum of an individual polymer embedded SWCNT (blue line). Spectrally filtered light (pink shaded bandpass) is directed to a photon coincidence setup (BS: beam splitter, APD: avalanche photodiode). (b,c) Corresponding plots of the second order photon correlation function under continuous wave excitation. Trace (b) was recorded with a 10 nm filter centered over the right half of the emission spectrum shown in (a) for a polystyrene embedded SWCNT. The photon bunching signature with a decay time of 3.7 ± 0.4 ns is a measure for the spectral diffusion time. Bunching is absent in trace (c) even with 1 nm narrow filtering for a pristine SWCNT bridging an air gap. Pump power is 200 µW for (a) and 1.8 mW for (b). All measurements were taken at 9 K.

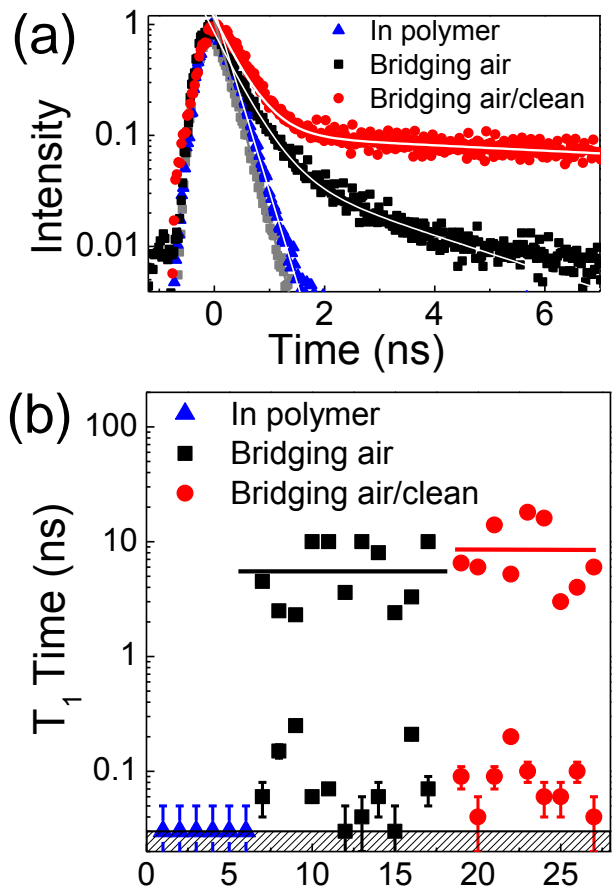

**Figure 4. Spontaneous emission time ($T_1$) recorded by TCSPC**. **(a)** Photoluminescence intensity as a function of delay time after pulsed excitation at 780 nm into a phonon sideband. The gray squares are the system response recorded for backreflected laser light and fit monoexponential to a decay time of 210 ps. All polymer embedded samples are nearly resolution limited (blue triangles) and fit monoexponential (white line) with a deconvolved decay time of 30±20 ps. Air-bridged SWCNTs from 10 min growth (black squares) and 2 min growth (red circles) fit biexponential (white lines) with a slower component of 30-200 ps and a faster component of several ns. **(b)** Overview of fitted $T_1$ times versus number of investigated SWCNT. Horizontal lines indicate average values of the slow component. Data are recorded at 9 K.

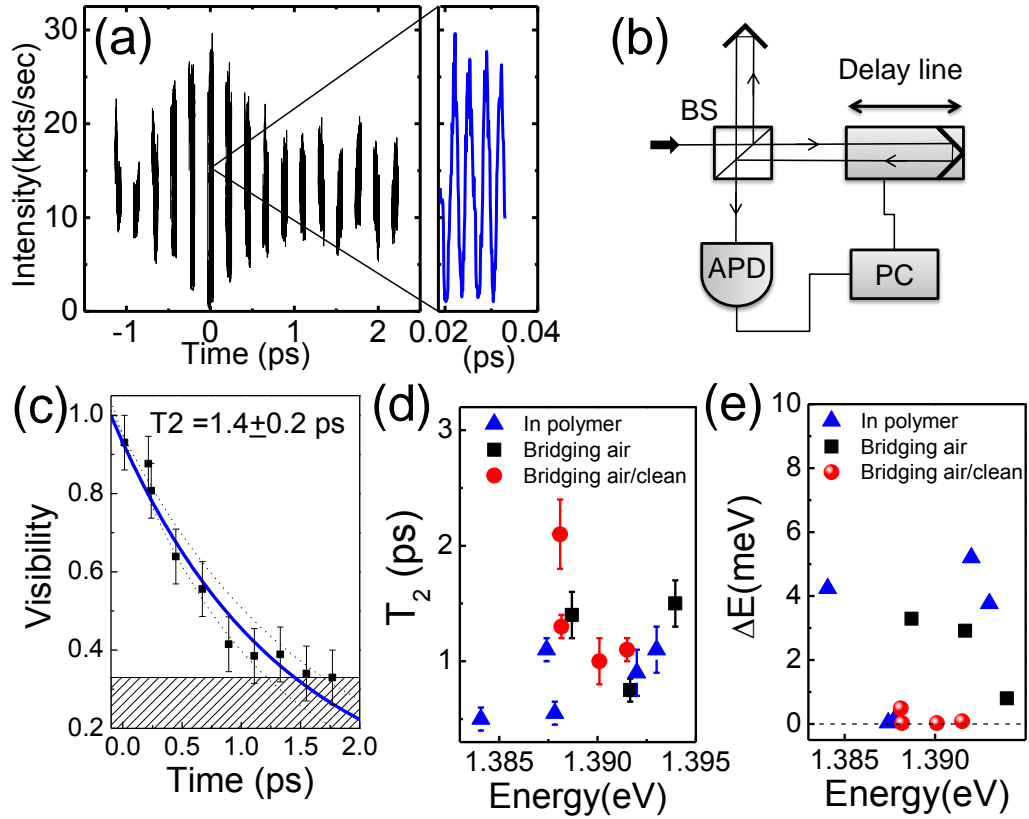

**Figure 5. Exciton dephasing time measurements of individual SWCNTs using a Michelson interferometer.** (a) Intensity as a function of delay time between the two beams in the interferometer. The right panel magnifies the pronounced fringe contrast near zero delay time. (b) Schematic of Michelson interferometer using retro mirrors on a delay line. BS: Beam splitter, APD: avalanche photo diode, PC: computer control. (c) Fringe visibility as a function of delay time recorded at 9K for a pristine SWCNT bridging an air gap (black dots). The blue solid line is a monoexponential decay fit. Dashed lines highlight the error bars. (d) Dephasing times for individual SWCNTs in polymer (blue triangles), bridging air for 10 min growth (black squares), and bridging air for 2 min growth (red circles). (e) Energy difference (ΔE) between SWCNTs linewidth determined from PL spectra and corresponding linewidth of the measured $T_2$ data from panel (d). Data are recorded at 9 K and pump power of about 900 µW.

# Prolonged spontaneous emission and dephasing of localized excitons in air-bridged carbon nanotubes – Supplementary Information

*Ibrahim Sarpkaya[1], Zhengyi Zhang[2], William Walden-Newman[1], Xuesi Wang[2], James Hone[2], Chee Wei Wong[2], and Stefan Strauf [1*]*

[1] Department of Physics, Stevens Institute of Technology, Hoboken, NJ 07030, USA

[2] Department of Mechanical Engineering, Columbia University, New York, NY 10027, USA

*Address correspondence to: strauf@stevens.edu

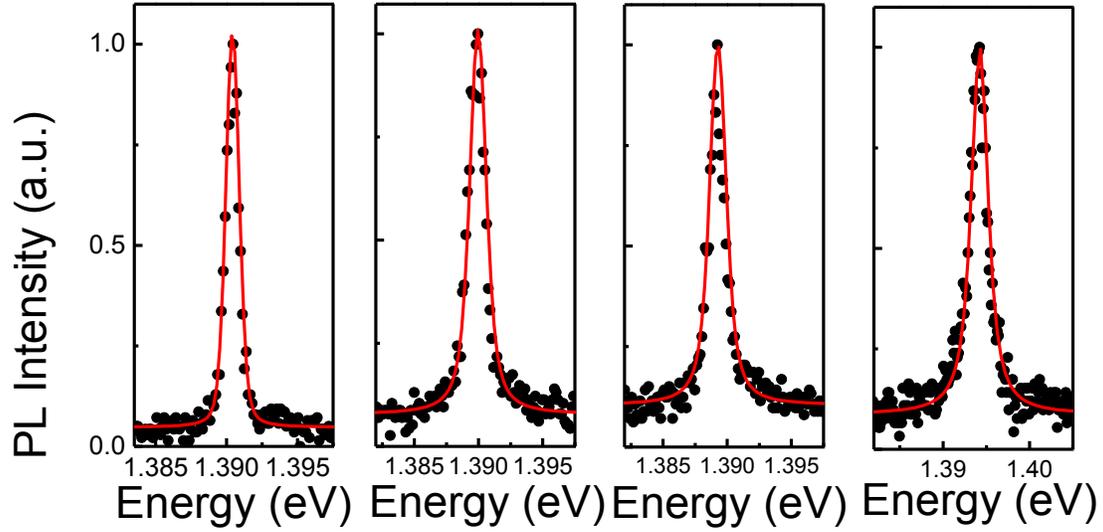

**Supplementary Figure S1: Normalized exciton emission spectra for individual air-bridged SWCNTs.** Four examples of the $E_{11}$ exciton transition of (6,4) chirality SWCNTs recorded at 9 K are shown, which fit well to Lorentzian lineshapes with no visible asymmetry at lower energies, in contrast to earlier reports. Previous studies reported asymmetric exciton lineshapes with linewidth of about 3.5 meV for surfactant dispersed SWCNTs grown by the CoMoCat technique, which were attributed to an intrinsic non-Markovian dephasing mechanism of excitons and acoustic phonons [29]. Such an asymmetry was also reported for air-bridged SWCNTs displaying rather large linewidth of 10 meV and attributed to the Van Hoove singularities in the density of states [30]. After the discovery that the optical emission in SWCNTs stems from excitons [2] it became clear that the lineshape is not related to the Van Hoove density of states. While contributions from exciton acoustic phonons to the lineshape are expected, we find that the lineshape is symmetric to a high degree and is with values down to 220 μeV also significantly narrower than previous work [2, 30]. We further quantified the residual degree of asymmetry on the lineshape by fitting a Doniach Sunjic lineshape to the data, yielding the asymmetry parameter α. For the four spectra we find rather small values of α=0.024, α=0.023, α=0.026, and α=0.01, respectively, suggesting for all intents and purposes, a symmetric lineshape.

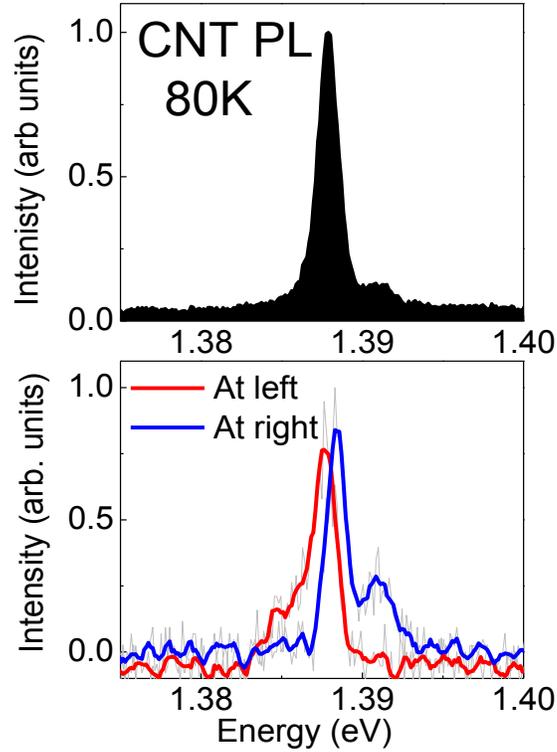

**Supplementary Figure S2: Illustration of the spectral filtering for the spectral diffusion measurements of air-bridged SWCNTs.** The top graph is the full PL spectrum recorded at 80 K from an individual SWCNT which has a FWHM of 1.7 meV (1.1 nm). The bottom graph shows the PL spectra after the light was coupled into an optical fiber, which was connected to a tunable 1 nm bandpass filter before reaching APDs, resulting in the filtered spectra with bandpass detuning to the left (red spectrum, sent to APD1) and to the right (blue spectrum, sent to APD2). We note that in this example the PL linewidth is about 20% narrower than for the experiment reported in Figure 3 in the main text (FWHM of 2.2 meV or 1.4 nm), but the spectra are still distinct. Therefore, the 1 nm bandpass filter creates a conditional probability at the detectors which should lead to the observation of bunching, if spectral diffusion contributes to the linewidth, and if the spectral diffusion time is slower than the timing jitter of the APDs.

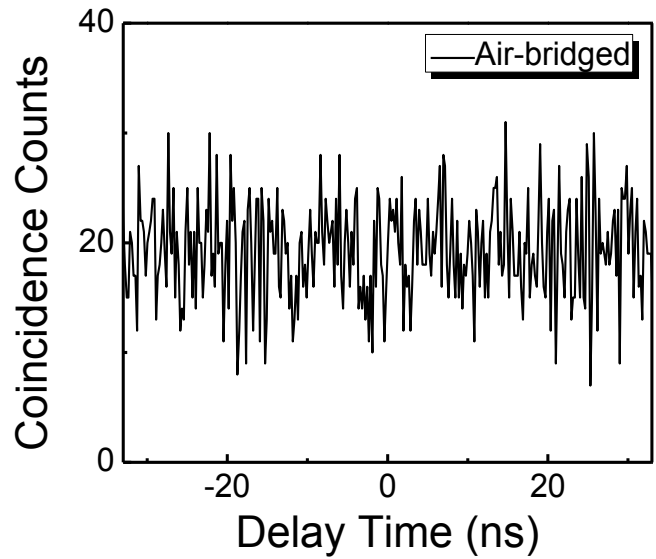

**Supplementary Figure S3: Photon correlation trace for the air-bridged SWCNT under continuous wave excitation.** Another reason why the correlation trace for the air-bridged tubes in Figure 3c of the main text might not show bunching from residual SD could be that it is masked by photon antibunching from localized quantum dot like excitons, in particular since some SWCNTs do show antibunching under pulsed excitation. In order to test this hypothesis we recorded photon correlation measurements for the same SWCNTs with a 40 nm broad bandpass filter that passes the entire PL spectrum. The result is a flat coincidence trace as shown, which was recorded for 1.5 hrs at 9 K**.** Even if the light emitted by the SWCNT would be antibunched it would be hard to be detectable since the APD timing jitter is about 210 ps and the exciton emission has a fast component of about 30 to 200 ps for the first order of magnitude (see Figure 3 in main text). We therefore conclude that antibunching does not mask the SD experiments.

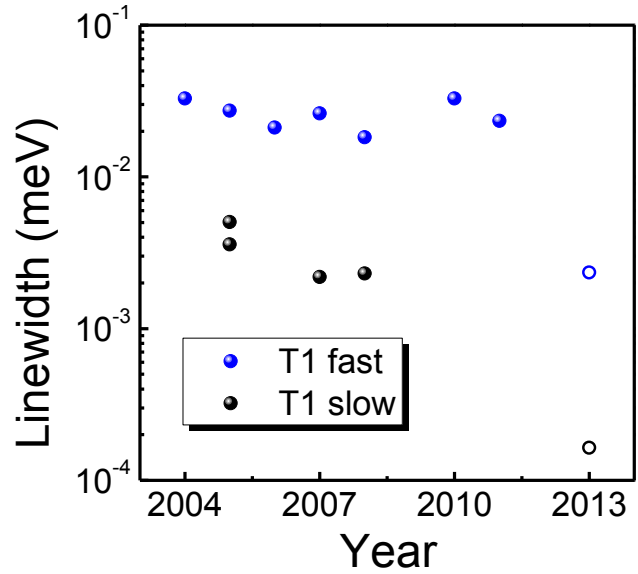

Supplementary Figure S4: Contribution to spectral linewidth from the spontaneous emission lifetime. In general, both $T_1$ and $T_2$ can contribute to the spectral linewidth via the well-known relation $\Gamma = \hbar/T_1 + 2\hbar/T_2$ where $T_2$ typically dominates the contribution to the linewidth. Our lifetime measurements in Figure 3 of the main text reveal however a fast component which is resolution limited, implying that $T_1$ could be much faster. To quantify this we compared the known values from 11 previously published papers (PRL, PRB, and Nano Letters) recorded with much higher temporal resolution and compared to our case and sorted them by publication year on the x-axis. Open data points in 2013 are this work. $T_1$-fast typically varies between 20 to 40 ps (but not faster) in surfactant dispersed CNTs. This corresponds to a spectral linewidth of only 20 to 40 µeV. While $T_1$ does not contribute more than 2% to the linewidth for polymer SWCNTs or for air-bridged SWCNTs from the 10 min growth, it contributes up to 20% for the narrowest linewidth of 220 µeV for the 2 min growth. We have therefore included the influence of $T_1$ for the linewidth values calculated from the interferometric dephasing time measurements and compared those corrected values to the spectral linewidth resulting in the ΔE values plotted in Figure 5e.

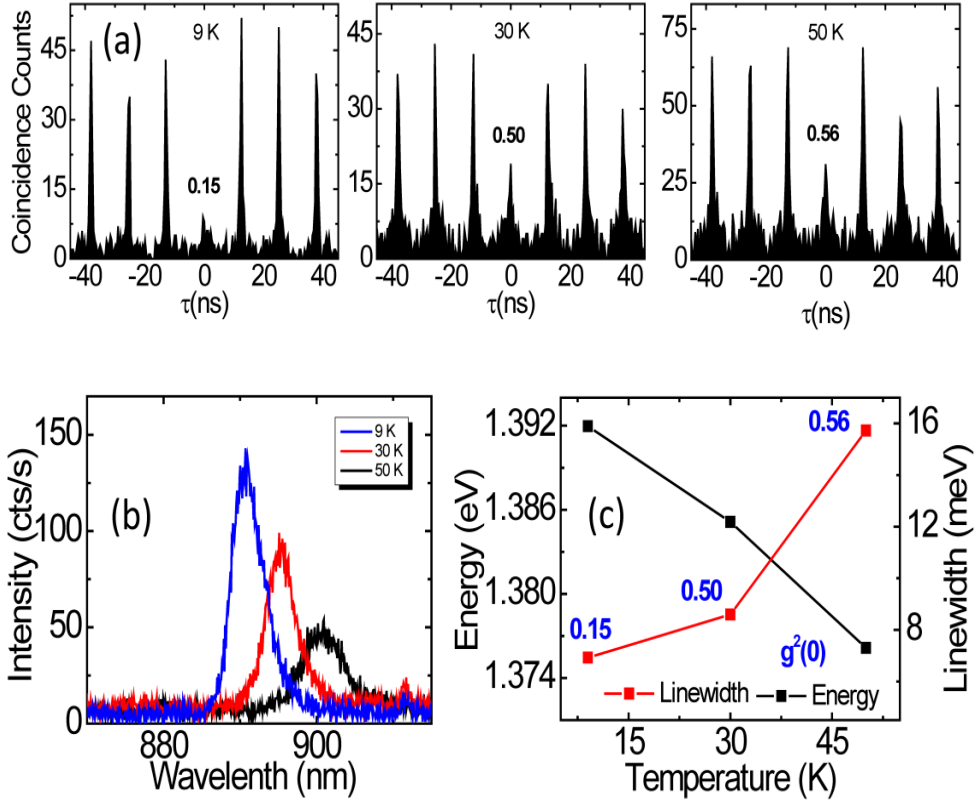

**Supplementary Figure S5: Temperature dependence of photon antibunching, photoluminescence, and linewidth for an individual polymer embedded SWCNT.** It is well known that quantum dot like states can form randomly along one-dimensional SWCNTs as is evident from Coulomb blockade effects in low-temperature electron transport experiments [40, 41]. These random confinement potentials along the tube axis can also give rise to spatial localization of optically excited excitons. This is most evident from recent low-temperature experiments showing pronounced photon antibunching of the spectrally filtered exciton emission [15, 16]. Since the exciton binding energy is about 400 meV larger than kT at room temperature one would expect quantum light signatures to survive up to room temperature. Another important factor is however the energetic depth of the localization potential which could give rise to exciton delocalization at elevated temperatures and thus a loss of quantum light signatures. In order to investigate the thermalization behavior of quantum-dot like excitons in SWCNTs we carried out photon antibunching studies at elevated temperatures using surfactant dispersed SWCNTs embedded in a

polystyrene cavity to enhance the light extraction [15]. Panel (a) shows the second order correlation function $g^{(2)}(\tau)$ for three different temperatures for a (6,4) chirality SWCNT, which has been recorded under pulsed excitation at 10 µW pump power. At 9 K we observe substantial antibunching with $g^{(2)}(0) = 0.15$ followed by a degrading of the single photon signature at 30 K with $g^{(2)}(0) = 0.5$, which further increases to $g^{(2)}(0) = 0.56$ at 50 K. Panel (b) shows the corresponding PL spectra which display a remarkable redshift of the peak energy (black squares) from 890 nm at 9 K to 900 nm at 50 K. This corresponds to a shift in emission energy of 15 meV as is shown in panel (c). The magnitude and direction of the shift is comparable to previous experiments with surfactant dispersed SWCNTs [29]. In that work, it was pointed out that the redshift cannot be reproduced in a non-Markovian model assuming coupling of excitons to the stretching mode phonons, and was thus attributed to environmental changes such as exciton delocalization. Our data set correlates the spectral redshift to a degradation of the single photon purity strongly supporting the idea of delocalization of quantum-dot like excitons at elevated temperatures in SWCNTs. We note that we have not been able to follow the thermalization behavior for this particular SWCNT up to higher temperatures since it exhibited at 80 K a sudden blue shift event and a strong diminishing of the PL intensity. Similar behavior was observed recently by Finnie and Levebre and attributed to molecular deposition and subsequent deep level formation under optical excitation [10].

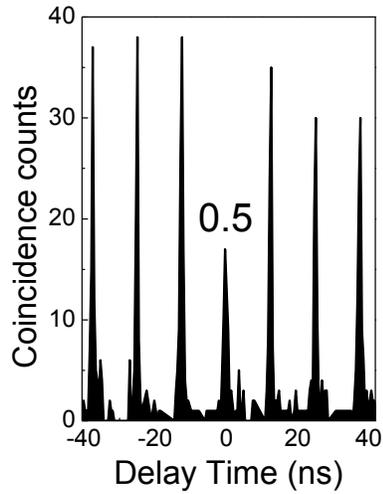

**Supplementary Figure S6: Photon antibunching of an air bridged SWCNT.** In order to show that localized quantum-dot like exciton states can also be present in air bridged SWCNTs at cryogenic temperatures, we recorded the second-order autocorrelation function for a (6,4) chirality SWCNT under pulsed optical excitation. The data are recorded at 9K for 1 hr and by pumping into a phonon sideband at 780 nm displaying nonclassical light emission with $g^{(2)}(0) = 0.5$. We remark that the majority of the investigated air-bridged SWCNTs do not display photon antibunching, while they all display a prolonged $T_1$ time of several ns in agreement with the absence of the PAIEI mechanism.